\documentclass[aoas]{imsart}

\RequirePackage{amsthm,amsmath,amsfonts,amssymb}
\RequirePackage[authoryear]{natbib}
\RequirePackage[colorlinks,citecolor=blue,urlcolor=blue]{hyperref}
\RequirePackage{graphicx}

\startlocaldefs
\theoremstyle{plain}

\newtheorem{theorem}{Theorem}[section]
\newtheorem{lemma}[theorem]{Lemma}
\theoremstyle{remark}

    \newtheorem{assumption}[theorem]{Assumption}

    \usepackage[ruled,vlined,linesnumbered]{algorithm2e}

    \usepackage{longtable} 
    \usepackage{tabularx} 
    \usepackage{relsize} 
    \newcolumntype{L}[1]{>{\raggedright\arraybackslash}p{#1}}
    \newcolumntype{C}[1]{>{\centering\arraybackslash}p{#1}}
    \newcolumntype{R}[1]{>{\raggedleft\arraybackslash}p{#1}}
    \usepackage{booktabs}
    \usepackage{rotating}
    \usepackage[para,online,flushleft]{threeparttable}

\endlocaldefs

\begin{document}

\begin{frontmatter}
\title{Non-convex SVM for cancer diagnosis based on morphologic features of tumor microenvironment}
\runtitle{Non-convex SVM on tumor microenvironment features}

\begin{aug}
\author[A]{\fnms{Sean}~\snm{Kent}\orcid{0000-0001-8697-9069}}
\and
\author[B]{\fnms{Menggang}~\snm{Yu}\ead[label=e2]{meyu@biostat.wisc.edu}\orcid{0000-0002-7904-3155}}

\address[A]{Department of Statistics,
University of Wisconsin -- Madison}

\address[B]{Department of Biostatistics and Medical Informatics,
University of Wisconsin -- Madison\printead[presep={,\ }]{e2}}
\end{aug}

\begin{abstract}
The surroundings of a cancerous tumor impact how it grows and develops in humans. New data from early breast cancer patients contains information on the collagen fibers surrounding the tumorous tissue---offering hope of finding additional biomarkers for diagnosis and prognosis---but poses two challenges for typical analysis. Each image section contains information on hundreds of fibers, and each tissue has multiple image sections contributing to a single prediction of tumor vs. non-tumor. This nested relationship of fibers within image spots within tissue samples requires a specialized analysis approach. 

We devise a novel support vector machine (SVM)-based predictive algorithm for this data structure. By treating the collection of fibers as a probability distribution, we can measure similarities between the collections through a flexible kernel approach. By assuming the relationship of tumor status between image sections and tissue samples, the constructed SVM problem is non-convex and traditional algorithms can not be applied. We propose two algorithms that exchange computational accuracy and efficiency to manage data of all sizes. The predictive performance of both algorithms is evaluated on the collagen fiber data set and additional simulation scenarios. We offer reproducible implementations of both algorithms of this approach in the R package \verb|mildsvm|. 
\end{abstract}


\end{frontmatter}

\section{Introduction} \label{intro}

The tumor microenvironment is the ecosystem that surrounds and interacts with a tumor. The composition of the tumor microenvironment includes the immune cells, stromal cells, blood vessels, and extracellular matrix (ECM) \citep{anderson_tumor_2020}. It is now widely accepted that the tumor microenvironment plays an important role in supporting tumor growth and progression \citep{anderson_tumor_2020}. Early in tumor growth, a dynamic and reciprocal relationship develops between cancer cells and components of the tumor microenvironment that supports cancer cell survival, local invasion, and metastatic dissemination \citep{baghban_tumor_2020}. Therefore, investigation of the tumor microenvironment can help discover noteworthy characteristics (e.g. biomarkers), leading to better cancer diagnosis and prognosis. Expanding literature on the tumor microenvironment has even identified new targets within it for therapeutic intervention \citep{bejarano_therapeutic_2021}.

In this article, we analyze morphological features of the ECM for breast cancer diagnosis. Breast tumor formation is associated with a stromal response, termed the desmoplastic reaction, characterized by amplified collagen matrix deposition and stromal cell recruitment and activation, thereby promoting tumor progression \citep{arendt_stroma_2010,zeltz_cancer-associated_2020}. Because increased cell numbers and increased collagen are sources of contrast within the mammogram, they are difficult to distinguish; traditional clinically proven methods, such as radiography and ultrasound imaging, do not have the resolution to differentiate the tumor from collagen at the cellular level. Recent imaging development, known as second harmonic generation (SHG) imaging \citep{chen_second_2012}, can accurately capture over 20 collagen features—including alignment, curvature, width, length, closeness to nearby fibers, and other features of ECM composition. These morphological features can be used as early biomarkers for cancer diagnosis and prognosis \citep{conklin_collagen_2018}. 

\begin{figure}[tbp]
    \centering\includegraphics[width=0.99\linewidth]{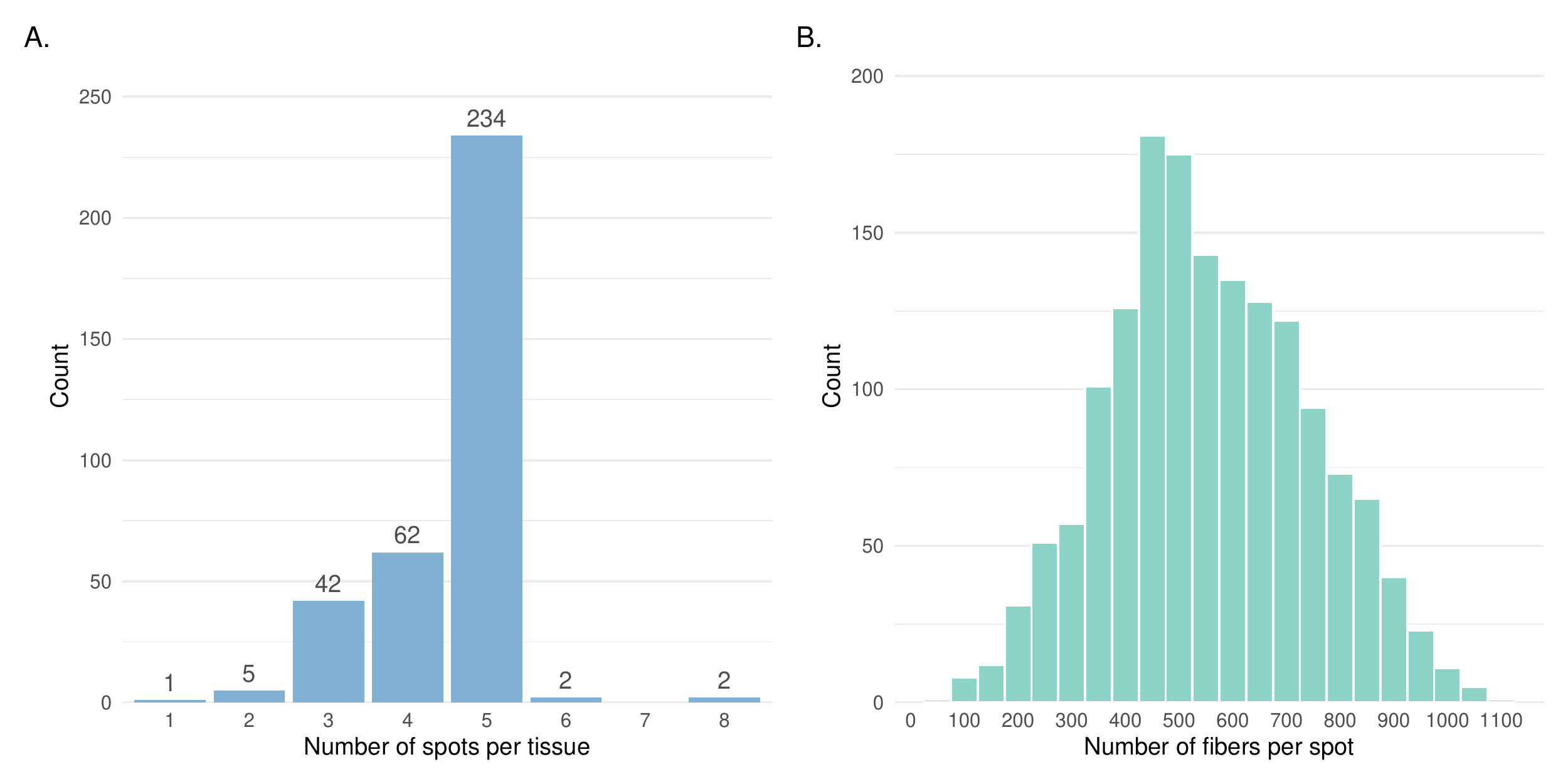}
    \caption{Number of spots and fibers in the data. \textbf{A.} Counts of the number of spots selected per tissue, with 5 spots being the most common. \textbf{B.} Counts of the number of collagen fibers per spot, with 450 being the most common.}
    \label{fig:hists}
\end{figure}

\begin{figure}[tbp]
    \centering\includegraphics[width=0.99\linewidth]{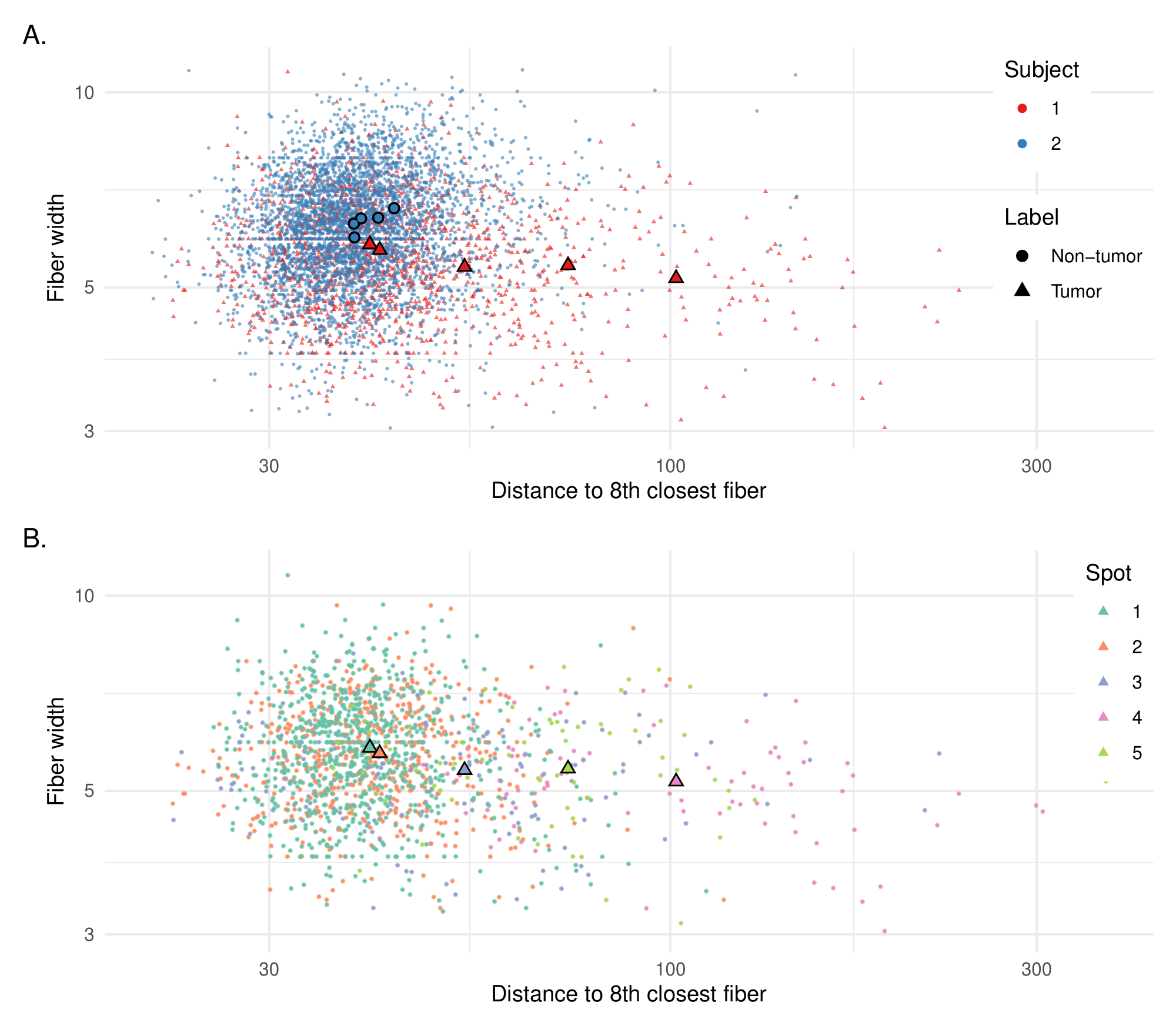}
    \caption{Example data from two subjects. Each small point represents the measurements on individual fibers within the tumor microenvironment, while large points (outlined in black) summarize the mean for each chosen spot. Both axes are on the log scale. \textbf{A.} Color distinguishes subjects while shape distinguishes whether the tissue sample was tumorous or not. We observe subject 1 with some spots close to the spots from subject 2, and other spots far from the cluster. \textbf{B.} Only subject 1 is shown. Color distinguishes the individual spots for this subject, showing distributional differences in both location and shape among the five spots chosen. }
    \label{fig:intro-example}
\end{figure}

A recent data set with these morphological features contains a nested structure that poses a few challenges. The data includes slides of breast tissue from ductal carcinoma in situ (DCIS) patients that are either tumor tissue or adjacent, normal tissue. For each tissue slide, between one and eight spots (median 5) are selected for SHG imaging (Figure~\ref{fig:hists}). Within each spot, SHG imaging captures features on hundreds of collagen fibers. Figure~\ref{fig:hists} shows detailed counts of the number of fibers per spot across the data set. An example of two subject slides, provided in Figure~\ref{fig:intro-example} for two informative features, illustrates two challenges in creating a tumor prediction algorithm. The first is how to summarize collagen relationships within each spot from the hundreds of fibers. Figure~\ref{fig:intro-example} shows that the mean and standard deviation of the distance to the eighth closest fiber may be predictors; for other features, correlations and quantiles may also be relevant. The second challenge is how to relate the summarized spot features to the prediction on the whole tissue slide where our training labels occur since each has many spots with information. 

We propose to tackle these challenges by connecting approaches from previous literature that address one of the two difficulties and constructing a non-convex support vector machine (SVM) learning algorithm. First, we capture the pooled information from the many fibers per spot by treating the collection as a sample from a multivariate probability distribution. This distribution summarizes the relationships \textit{between} collagen fibers rather than looking individually \textit{within} them. We leverage growing literature on learning from probability distributions in works such as \citet{muandet_learning_2012,muandet_kernel_2017,lin_learning_2013}, which is part of the broader category of functional data analysis \citep{ramsay_functional_2006}. Second, we assume that for each tumor slide, at least one of the spots has signs of cancer. Conversely, among non-tumor slides, no spots have tumor features. This assumption creates weak supervision between the labels at the spot level and the predictions at the tissue sample level \citep{campanella_clinical-grade_2019,li_convex_2013}. Our approach combines both ideas and minimizes a hinge loss function on the data while maximizing the margin of error for classification, resulting in a non-convex type of SVM. 

Our proposed methodology introduces two algorithmic solutions to the non-convex SVM. One heuristically moves between selecting important spots and then solving a reduced problem on the selected spots. This approach has roots in several optimization approaches, including the Expectation-Maximization algorithm \citep{moon_expectation-maximization_1996}, the Convex Concave Procedure \citep{yuille_concave-convex_2003}, and a similar non-convex SVM for weakly supervised data \citep{andrews_support_2003}. The other approach solves the problem directly by modifying its form and using efficient algorithms for problems involving integer variables \citep{gurobi_optimization_llc_mixed-integer_2021}. In doing so, we contribute new ideas for approximating non-linear kernels when learning from data with probability distributions. Both methods are applied to the motivating data on tumor microenvironment features, and they perform favorably without requiring the researcher to select the meaningful collagen features \textit{a priori}. We also provide an R package to solve the non-convex SVM problem for data with a similar structure. 

The remainder of this paper contains the following: a detailed account of our methodology---including a formalized data set structure, non-convex SVM derivations, and two computational algorithms---in Section~\ref{methods}; predictive performance on simulated data sets in Section~\ref{simulations}; an empirical evaluation on the motivating collagen feature data set in Section~\ref{analysis}; and a discussion of methodological considerations, implications, and future research directions in Section~\ref{discussion}.

\section{Methodology} \label{methods}

In this section, we formulate a predictive model from data structured with weakly-supervised outcome labels and distributional instances---as seen in the motivating collagen fiber data set. We will transform the joint goal of minimizing training loss and maximizing the error margin into a non-convex SVM optimization problem. We then develop two separate algorithms to solve the SVM problem: one with a heuristic approach iteratively solving a convex sub-problem and another directly solving the problem after a feature map approximation for probability distributions. 

\subsection{Data set structure} 

Consider a data set with $n$ distributional instances. Let $\mathcal{P}$ denote the set of all probability measures $\mathbb{P}$ over the data domain $\mathcal{X} \subseteq \mathbb{R}^d$ for $d \in \mathbb{N}$. Formally, we say that each instance $i$ has information contained by a probability measure $\mathbb{P}_i$ on the data domain $\mathcal{X}$. In practice, this distribution could manifest in the data as a sample of size $r_i$ drawn from $\mathbb{P}_i$, denoted as $\hat{\mathbb{P}}_i = \{x_{i,j}\}_{j=1}^{r_i}$, where $x_{i,j} \in \mathcal{X}$. Collectively, we can refer to this data set as $\{\mathbb{P}_i\}_{i=1}^n$ or $\{\hat{\mathbb{P}}_i\}_{i=1}^n$. In our motivating data set, each distributional instances is a spot with samples from the individual collagen fibers. 

Next, we consider weakly-supervised outcome labels within the data set. Let $\mathcal{I}$ represent the set of all top-level groups in the data, with each group $I \in \mathcal{I}$ having a corresponding label of interest given by $Y_I \in \{ -1, 1 \}$. Each top-level group is a collection of spots $i \in I$ and can thus be thought of as an index set.\footnote{The index sets $I \in \mathcal{I}$ together create a partition of the spots $\{1,2,\dots,n \}$ into mutually exclusive groups that include all spots: $\bigcup_{I \in \mathcal{I}} I = \{1, \dots, n\}$ and $I \cap J = \emptyset$ for all pairs $I, J \in \mathcal{I}$.} The data set information for each top-level group is denoted by $B_I = \{ \mathbb{P}_i : i \in I\}$. Each probability distributions $\mathbb{P}_i$ within $I$ has a potentially \textit{unobserved} binary label $y_i \in \{-1, 1\}$, and, collectively, they contribute to the top-level label $Y_I$. This nested structure results in a \textit{weak supervision} between the spots and the outcome, and the term has been coined in previous works \citep{campanella_clinical-grade_2019,li_convex_2013}. In our data, the top-level groups correspond to tissue slides, and labels refer to tumor (+1) and non-tumor (-1).  

\subsection{Non-convex SVM formulation} \label{sec:svm}

Our goal is to find a function $F$ such that $F(B_I) \approx Y_I$ using the collective information from $\{ \mathbb{P}_i : i \in I\}$.\footnote{We use $\approx$ to indicate closeness with respect to some loss function}  First, we assume that the spot level outcomes $y_i$ relate to the slide-level group outcomes $Y_I$ as follows: a positive slide label $(Y_I = 1)$ is observed if and only if at least one spot label in it is positive $(\exists i \in I : y_i = 1)$. Mathematically, this is equivalent to the following assumption:
\begin{assumption}
\label{a:mil}
$Y_I = \max_{i \in I} y_i \quad \forall I \in \mathcal{I}$.
\end{assumption}
Assumption~\ref{a:mil} works well for the motivating data set on collagen fibers using tumor for the positive label. We assume that if a single spot on the tissue sample is tumorous, the specimen is tumorous. Conversely, non-tumor samples must have all spots with non-tumorous features. This assumption could be violated if no tumorous areas were selected as spots in a tumor tissue sample. However, the procedure for determining spots had experts focus on tumorous areas, making this violation unlikely in practice.

Working at the spot level, we suppose that a linear form is sufficient to classify labels after a possible non-linear transformation $\phi$, so that $f(\mathbb{P}_i) = \langle \boldsymbol{w}, \phi(\mathbb{P}_i) \rangle + b$ and $f(\mathbb{P}_i) \approx y_i$. For the moment, consider $\boldsymbol{w}$ and $\phi$ as finite dimensional. Then following Assumption~\ref{a:mil}, a natural form for $F$ is $F(B_I) = \max_{i \in I} f(\mathbb{P}_i)$. The hyperplane $F(B_I) = 0$ divides the slide-level groups into two regions which are classified as positive and negative, and in a similar way $f(\mathbb{P}_i) = 0$ divides the instances into positive and negative regions. 

Following an SVM approach \citep{burges_tutorial_1998} we seek to 1) minimize the hinge loss function $\ell(y, x) = \max(0, 1 - y x)$ between labels $Y_I$ and predictions $F(B_I)$ for all $I$ and 2) maximize the margin around the hyperplane created by $f(\mathbb{P}_i) = 0$, which is equivalent to minimizing $\left\lVert \boldsymbol{w} \right\rVert^2$. Hinge loss is used because if $\ell\left(Y_I, \max_{i \in I} \langle \boldsymbol{w}, \phi(\mathbb{P}_i) \rangle + b\right) = 0$, then $Y_I \max_{i \in I} \left( \langle \boldsymbol{w}, \phi(\mathbb{P}_i) \rangle + b \right) \geq 1$, i.e. group labels fall outside the margin of the classifier and are correctly classified; otherwise, the hinge loss increases proportionally to the distance from that margin. Larger margins are beneficial because they offer the hope of strong performance on unseen data. We balance these two objectives with a tuneable hyper-parameter $C$: 
\begin{equation*}
    \min_{\boldsymbol{w}, b} \left\{ \frac{1}{2} \left\lVert \boldsymbol{w} \right\rVert^2 + C \sum_{I \in \mathcal{I}} \ell\left(Y_I, \max_{i \in I} \langle \boldsymbol{w}, \phi(\mathbb{P}_i) \rangle + b\right) \right\}
\end{equation*}

We can re-write this objective in a more common form by using auxiliary variables $\boldsymbol{\xi} = \left[\xi_I \right]_{I \in \mathcal{I}}$ defined by $\xi_I = \ell\left(Y_I, \max_{i \in I} \langle \boldsymbol{w}, \phi(\mathbb{P}_i) \rangle + b\right)$. From the definition of $\ell$, $\xi_I$ is the smallest non-negative number such that $Y_I \max_{i \in I} \left( \langle \boldsymbol{w}, \phi(\mathbb{P}_i) \rangle + b \right) \geq 1 - \xi_I$, and so we have the following optimization problem:
\begin{equation}
\label{MI-SMM primal}    
\begin{split}
    & \min_{\boldsymbol{w}, b, \boldsymbol{\xi} \geq 0}  \frac{1}{2} \left\lVert \boldsymbol{w} \right\rVert^2 + C \sum_{I \in \mathcal{I}} \xi_I
    \\
    \textrm{s.t. \quad} & Y_I \max_{i \in I} \left( \langle \boldsymbol{w}, \phi(\mathbb{P}_i) \rangle + b \right) \geq 1 - \xi_I, \enskip \forall I \in \mathcal{I}
\end{split}
\end{equation}
where $\boldsymbol{\xi} \geq 0$ indicates that $\xi_I \geq 0 \enskip \forall I$. We refer to \eqref{MI-SMM primal} as the primal problem. It is a non-convex mixed-integer quadratic programming (MIQP) problem. The mixed-integer nature can be revealed by explicitly considering which $i \in I$ yields the maximum value of $\langle \boldsymbol{w}, \phi(\mathbb{P}_i) \rangle + b$ for each of the slide-level groups in the constraint of \eqref{MI-SMM primal}. 

Related research has studied non-convex SVMs under several different contexts. One substantial branch added a non-convex penalty to the hinge loss to induce sparsity for variable selection \citep{zhang_gene_2006,laporte_nonconvex_2013,zhang_variable_2016,liu_global_2016,guan_efficient_2020}. Another introduced non-convexity through the ramp loss to reduce the influence of outliers \citep{ertekin_nonconvex_2010,zhao_two-stage_2021}. Our approach in Problem \eqref{MI-SMM primal} differs because non-convexity results from the $\max$ operator  in the constraint, instead of from a change in the loss function. \cite{andrews_support_2003} arrive at a similar formulation when considering standard vector data instead of distributional instances. In both the semi- and weakly-supervised settings, \cite{li_convex_2013} explore a more general non-convex, MIQP SVM problem through a convex relaxation approach. 

\subsection{A heuristic algorithm from dual SVM problem} \label{sec:heuristic}

 To calculate predictive models from data structured with weakly-supervised outcome labels and distributional instances, we need algorithms that can solve the optimization problem in \eqref{MI-SMM primal}. Our first algorithm, which we will call \textsc{mi-smm~(heuristic)}, relies on finding the dual to a convex sub-problem of \eqref{MI-SMM primal} and iteratively solving that problem using an appropriate kernel function on distributional instances. 
 
\subsubsection{The convex sub-problem and dual SVM} \label{sec:dual-svm}

While Problem \eqref{MI-SMM primal} has an intuitive setup, it is often inconvenient to work with.  If $\boldsymbol{w}$ and $\phi$ are infinite-dimensional or the form of $\phi$ is unknown, solving \eqref{MI-SMM primal} is intractable. One way of resolving the intractability is by using Wolfe duality theory to find an equivalent problem on which the ``kernel trick'' can be applied \citep{boyd_5_2004}. However, many of the guarantees within duality theory do not apply to non-convex problems. 

One solution is to find a convex sub-problem within \eqref{MI-SMM primal}. Consider a function $s$ that selects one spot $i$ from each slide-level group $I$ having positive label $Y_I = 1$, so that $s(I) = i$. An optimal $s$ will choose $s(I) = \arg\max_{i \in I} \left( \langle \boldsymbol{w}, \phi(\mathbb{P}_i) \rangle + b \right)$. For any fixed (potentially non-optimal) $s$, we can solve the convex SVM problem:
\begin{equation}
\label{MI-SMM sel}    
\begin{split}
    & \min_{\boldsymbol{w}, b, \boldsymbol{\xi} \geq 0}  \frac{1}{2} \left\lVert \boldsymbol{w} \right\rVert^2 + C \sum_{I \in \mathcal{I}} \xi_I
    \\
    \textrm{s.t. \quad} & Y_I \left( \langle \boldsymbol{w}, \phi(\mathbb{P}_{s(I)}) \rangle + b \right) \geq 1 - \xi_I, \enskip \forall I \in \mathcal{I}
\end{split}
\end{equation}

On the convex SVM problem \eqref{MI-SMM sel}, we can follow standard duality theory by calculating a Lagrangian function and evaluating the Karush–Kuhn–Tucker conditions \citep{boyd_5_2004} to find the dual problem, which has a similar form as that of \citet{andrews_support_2003}:
\begin{equation}
\label{MI-SMM dual}
\begin{split}
    \max_{\boldsymbol{\alpha}} &\sum_{i \in E} \alpha_i - \frac{1}{2} \sum_{i \in E} \sum_{j \in E} \alpha_i \alpha_j Y_{B(i)} Y_{B(j)} 
    \langle \phi(\mathbb{P}_i), \phi(\mathbb{P}_j) \rangle
    \\
    \textrm{s.t. \quad} \forall I \in \mathcal{I} \text{ : }& Y_I = +1 \text{ and } 0 \leq \alpha_i \leq C, \enskip \forall i \in I, 
    \\
    \text{or }& Y_I = -1 \text{ and } 0 \leq \sum_{i \in I} \alpha_i \leq C, 
    \\
    & \sum_{i \in E} \alpha_i Y_{B(i)} = 0, \quad \text{and} \quad \alpha_i \geq 0, \enskip \forall i \in E
\end{split}
\end{equation}
where $E = \left( \bigcup_{I : Y_I = -1} \{i \in I\} \right) \cup \left( \bigcup_{I : Y_I = +1} s(I) \right)$ is the effective set of indexes consisting of all negative spots and the selected spot from each positive group, and where $B(i)$ is the slide associated with spot $i$ (i.e. $B(i) = I$ for $i \in I$). Problem \eqref{MI-SMM dual} is a convex quadratic program that is solvable by standard QP solvers or by chunking algorithms like the sequential minimal optimization (SMO) algorithm \citep{platt_sequential_1998,boser_training_1992}. 

Similar to the standard SVM case, we construct the classifier based on $\alpha_i$ as $f(\mathbb{P}) = \sum_{i \in E} \alpha_i Y_{B(i)}  \langle \phi(\mathbb{P}), \phi(\mathbb{P}_i) \rangle + b$.  Given the final $h$, we compute predictions on new slides $B_{I'}$ as $\textrm{sign} \left( \max_{i \in I'} h(\mathbb{P}_i) > 0 \right)$, 
although the threshold can also be modified away from 0 to tailor the sensitivity or specificity of the classifier. 

The constant $b$ is determined by solving the following equation for a given spot $i$, or by taking the average over all such equations, as has been suggested for SVM \citep{burges_tutorial_1998}: 
\begin{equation}
\label{b}
\begin{gathered}
    Y_{B(i)} \left( \sum_{j \in E} \alpha_j Y_{B(j)} K(\mathbb{P}_i, \mathbb{P}_j) + b \right)= 1 
    \\
    \textrm{s.t. }  0 < \alpha_i < C \enskip\text{if } Y_{B(i)} = +1 \quad
    \text{or}\quad  0 < \sum_{j \in B(i)} \alpha_j < C \enskip\text{if } Y_{B(i)} = -1
\end{gathered}
\end{equation}

Problem \eqref{MI-SMM primal} is equivalent to the minimum of all choices of selection functions $s$ of Problem \eqref{MI-SMM dual}. The last step to solving the intractability in \eqref{MI-SMM primal} when $w$ and $\phi$ are infinite-dimensional is to utilize the ``kernel trick'' by replacing $\langle \phi(\mathbb{P}), \phi(\mathbb{Q}) \rangle$ with a kernel function $K(\mathbb{P}, \mathbb{Q})$. While kernels on vector data are commonly seen, a few works consider kernels appropriate for distributional instances. The following sub-section describes the support measure machine (SMM) kernel, an approach by \citet{muandet_learning_2012} and \cite{muandet_kernel_2017}, that offers flexibility in choosing the underlying non-linearity similar to kernels on vector data.

\subsubsection{SMM kernel} \label{sec:smm}

The approach from \cite{muandet_learning_2012} and \cite{muandet_kernel_2017} involves a family of kernels defined on probability distributions, which they use to learn a classifier from the SVM dual problem. 

Given a reproducing kernel $k : \mathcal{X} \times \mathcal{X} \to \mathbb{R}$ associated with a reproducing kernel Hilbert space (RKHS) $\mathcal{H}$, we can define $K : \mathcal{P} \times \mathcal{P} \to \mathbb{R}$ by 
\begin{equation*}
\label{smm-kernel}
    K(\mathbb{P}_i, \mathbb{P}_j) = \mathbb{E}_{X\sim \mathbb{P}_i, Y \sim \mathbb{P}_j} \left[ k(X, Y) \right] = \int \int k(x,y) d\mathbb{P}_i(X) d\mathbb{P}_j(Y) 
\end{equation*}
They show under standard assumptions that $K$ is a positive definite kernel on $\mathcal{P}$ and that $K$ works as an inner-product of so-called ``mean maps''. Let $\mu : \mathcal{P} \to \mathcal{H}$ denote the mean map: $\mu(\mathbb{P}) = \int_\mathcal{X} k(x, \cdot) d \mathbb{P}(x)$.  It follows that $K(\mathbb{P}, \mathbb{Q}) = \langle \mu(\mathbb{P}), \mu(\mathbb{Q}) \rangle_{\mathcal{H}}$.  

In some cases, this kernel can be computed in closed form \citep{muandet_learning_2012}.  When this is not possible, we can approximate $K(\mathbb{P}_i, \mathbb{P}_j)$ when the distributions manifest as the empirical distributions $\hat{\mathbb{P}}_i$ and $\hat{\mathbb{P}}_j$ of random samples $\{x_{i,l}\}_{l=1}^{r_i}$ and $\{z_{j,m}\}_{m=1}^{r_j}$, respectively:
\begin{equation*}
\label{smm-ekernel}
    K_\textrm{emp}(\hat{\mathbb{P}}_i, \hat{\mathbb{P}}_j) 
    = \frac{1}{r_i \cdot r_j} \sum_{l=1}^{r_i} \sum_{m=1}^{r_j} k(x_{i,l}, z_{j,m})
\end{equation*}

Once the kernel $K(\mathbb{P}_i, \mathbb{P}_j)$ is computed for all combinations $i,j$, it can be plugged into an SVM dual problem such as \eqref{MI-SMM dual}.  

\subsubsection{Algorithm} \label{sec:heur-alg}

Since the dual problem \eqref{MI-SMM dual} relies on a fixed selection function $s$, we can use it in a heuristic algorithm that updates the model at each step based on a solution to the convex SVM problem \eqref{MI-SMM dual} and subsequently updates the selection function $s$. This idea is similar to that of the Expectation-Maximization (E-M) algorithm \citep{moon_expectation-maximization_1996}, the Convex Concave Procedure \citep{yuille_concave-convex_2003}, and a non-convex SVM with weak labels studied previously \citep{andrews_support_2003}. We describe the algorithm in detail in Algorithm~\ref{alg2}.  

\begin{algorithm}[tbp]
\label{alg2}
\caption{\textsc{mi-smm~(heuristic)}}
\DontPrintSemicolon
\SetAlgoLined

\SetKw{Initialize}{initialize}
    \KwIn{training data $\left\{ Y_{B(i)}, \mathbb{P}_i \right\}_{i=1}^{n}$, embedding kernel $k$.}
    \BlankLine
    \Initialize{ $s(I)$ as a random selector function, $\forall I$ such that $Y_I = 1$ }
    \;
    Pre-compute individual kernel matrix entries $K(\mathbb{P}_i, \mathbb{P}_j)$ for all combinations $i, j$ in the data\; 
    \While{(selector variables $s(I)$ have changed)}{
        $E \leftarrow \left( \bigcup_{I : Y_I = -1} \{i \in I\} \right) \cup \left( \bigcup_{I : Y_I = +1} s(I) \right)$\;
        Solve for $\{\alpha_i\}_{i \in E}$ in Problem \eqref{MI-SMM dual} based on the instances within $E$\;
        Compute $b$ from \eqref{b}, or an average over all eligible equations \;
        $h(\mathbb{P}_i) \leftarrow \sum_{j \in E} \alpha_j Y_{B(j)} K(\mathbb{P}_i, \mathbb{P}_j) + b$ for all instances $\mathbb{P}_i$ in positive groups\;
        $s(I) \leftarrow \arg\max_{i \in I} h(\mathbb{P}_i)\; \forall I$ such that $Y_I = 1$\;
    }
    \KwRet SMM classifier $h(\mathbb{P}) = \sum_{j \in E} \alpha_j Y_{B(j)} K(\mathbb{P}, \mathbb{P}_j) + b$
\end{algorithm} 

This algorithm can be improved by only computing entries of $K(\mathbb{P}_i, \mathbb{P}_j)$ as needed in the set $E$, which will increase the speed slightly.  Additionally, one could perform Algorithm~\ref{alg2} over numerous random initial selections $s$ and combine the outputs by ensembling or choosing the highest performing version from a hold-out data set. It is also common to limit the number of times selector variables can change to avoid getting stuck in the while block forever. 

Even though Algorithm~\ref{alg2} and similar approaches empirically show good performance, there is no guarantee that it will find the globally optimal solution to \eqref{MI-SMM primal} \citep{andrews_support_2003,yuille_concave-convex_2003}. The algorithm can get stuck in local minimum, and it is difficult to know when this occurs. 

\subsection{A direct algorithm from MIQP formulation} \label{sec:miqp}

As an alternative to the heuristic approach in Section~\ref{sec:heuristic}, we can solve for the MI-SMM classifier in a more direct fashion.  We call this approach \textsc{mi-smm (miqp)}. For simplicity of construction, assume first that the form of $\phi$ in Problem \eqref{MI-SMM primal} maps explicitly to a real-valued feature vector: $\phi(\mathbb{P}_i) = z_i \in \mathbb{R}^p$. We define index sets of the spots based on outcome labels as $\mathcal{I}_+ = \{I : Y_I = 1\}$ and $\mathcal{I}_- = \{I : Y_I = -1\}$.  Using a well-known optimization trick (Lemma provided in the Supplementary Materials), Problem \eqref{MI-SMM primal} is equivalent to
\begin{equation}
\label{MI-SMM MIP}
\begin{split}
    & \min_{\boldsymbol{w}, b, \boldsymbol{\xi} \geq 0, \boldsymbol{\zeta} \in \{0, 1\}}  \frac{1}{2} \left\lVert \boldsymbol{w} \right\rVert^2 + C \sum_{I} \xi_I
    \\
    \textrm{s.t. \quad} & - \langle \boldsymbol{w}, z_i \rangle - b \geq 1 - \xi_I, \enskip \forall i \in I, \enskip \forall I \in \mathcal{I}_-
    \\ 
    & + \langle \boldsymbol{w}, z_i \rangle + b \geq 1 - \xi_I - L \cdot \zeta_{I, i}, \enskip \forall i \in I,  \enskip \forall I \in \mathcal{I}_+
    \\
    &\sum_{i \in I} \zeta_{I,i} \leq |I| - 1, \enskip \forall I \in \mathcal{I}_+ 
\end{split}
\end{equation}
provided that $L$ is a sufficiently large scalar.  In the above formulation, $|I|$ is the cardinality of $I$, $\boldsymbol{\zeta} = \left[\zeta_{I,i} \right]_{I \in \mathcal{I}_{+}, i \in I}$, and $\boldsymbol{\zeta} \in \{0, 1\}$ means $\zeta_{I,i} \in \{0,1\} \enskip \forall I \in \mathcal{I}_{+}$, $i \in I$.  This unpacks the $\max$ constraint for positively labeled slides through the introduction of binary variables $\zeta_{I,i}$.  For a given slide-level group $I$, not all of $\zeta_{I,i}$ can be 1, which ensures that $\langle \boldsymbol{w}, z_i \rangle + b \geq 1 - \xi_I$ for at least one spot $i$ in each slide.  

\subsubsection{Feature map approximation} \label{sec:fm-approx}

An unfortunate aspect of the formulation in \eqref{MI-SMM MIP} is that it is still non-convex, leaving out the option of using duality theory to find an equivalent formulation that directly uses the kernel $K(\mathbb{P}, \mathbb{Q})$, as was done in Section~\ref{sec:dual-svm}.  When the form of $\phi$ does not map to a real-valued feature vector, Problem \eqref{MI-SMM MIP} is intractable.  To get around this challenge, we leverage the theory developed on feature map approximations \citep{hutchison_mercers_2006,rahimi_random_2008,vedaldi_efficient_2012} to find a real-valued function $\tilde{\phi}$ that approximates a kernel $k$ through inner products:
\begin{equation*}
    k(x, z) \approx \langle \tilde{\phi}(x), \tilde{\phi}(z) \rangle \quad \forall x, z \in \mathcal{X}
\end{equation*}

There is little discussion of feature map approximations on probability distributions. Lemma \ref{lemma-fm} below provides some intuition for the SMM kernel in a special case.  The empirical kernel $K_{emp}$ summarizes the underlying feature map by taking the mean of that map over the distribution samples, and non-linearity is introduced by the embedding kernel $k$ and its feature map $\phi$. We provide a short proof in the Supplementary Material since  \citet{muandet_learning_2012} do not describe this result.

\begin{lemma}
\label{lemma-fm}
    Let $k : X \times X \to \mathbb{R}$ from a RKHS admit a feature map $\phi : X \to F \subset \mathbb{R}^p$ so that $k(x,z) = \langle \phi(x), \phi(z) \rangle$.  Then 
    \begin{equation*}
        K_{\textrm{emp}} (\hat{\mathbb{P}}_i, \hat{\mathbb{P}}_j) = \langle \bar{\phi}(\hat{\mathbb{P}}_i), \bar{\phi}(\hat{\mathbb{P}}_j) \rangle
    \end{equation*}
    where $\bar{\phi}(\hat{\mathbb{P}}_i) = \frac{1}{r_i} \sum_{l = 1}^{r_i} \phi(x_{i,l})$ and $\bar{\phi}(\hat{\mathbb{P}}_j) = \frac{1}{r_j} \sum_{m = 1}^{r_j} \phi(x_{j,m})$ are the empirical means of the random samples evaluated on the feature map.  
\end{lemma}

Give a feature map approximation $\tilde{\phi}$ for the underlying embedding kernel $k$, Lemma \ref{lemma-fm} provides a natural extension to a feature map approximation for $K$: 
\begin{equation*}
    K_{emp}(\hat{\mathbb{P}}_i, \hat{\mathbb{P}}_j) \approx 
    \left\langle \frac{1}{r_i} \sum_{l = 1}^{r_i} \tilde{\phi}(x_{i,l}), \frac{1}{r_j} \sum_{m = 1}^{r_j} \tilde{\phi}(x_{j,m}) \right\rangle,
\end{equation*}
for the random samples $\{x_{i,l} \}_{l=1}^{r_i}$ and $\{x_{j,m} \}_{m=1}^{r_j}$ from empirical distributions $\hat{\mathbb{P}}_i$ and $\hat{\mathbb{P}}_j$, respectively. 
Computationally, this is much faster than approximating $K$ as $\frac{1}{r_i \cdot r_j} \sum_{l = 1}^{r_i} \sum_{m = 1}^{r_j} k(x_{i,l}, x_{j,m})$, since it requires calculating $k(x_{i,l}, x_{j,m})$ from $\langle \tilde{\phi}(x_{i,l}), \tilde{\phi}(x_{j,m}) \rangle$ for all combinations of samples. 

In this paper, we use the Nystr\"{o}m feature map \citep{williams_using_2001} to approximate the embedding kernel $k$ because of its simplicity and excellent performance.  \citet{yang_nystrom_2012} provide a detailed description and intuition for this feature map, and \citet{chatalic_nystrom_2022} have recently applied Nystr\"{o}m methods to kernels on probability distributions. We modify the Nystr\"{o}m algorithm by taking a stratified sub-sample from the concatenation of all samples $\left\{ \{x_{i,l}\}_{l=1}^{r_i} \right\}_{i=1}^{n}$ instead of a simple sub-sample, ensuring each slide-level group is represented almost uniformly (details follow in Algorithm~\ref{alg3}).  

\subsubsection{Algorithm} \label{sec:alg2}

Algorithm \ref{alg3} connects the Nystr\"{o}m feature map procedure, SMM kernel approximation, and MIQP from Problem \eqref{MI-SMM MIP} into the direct approach, \textsc{mi-smm (miqp)}.  It requires a MIQP solver and two integer constants, $m_1$ and $m_2$, for the kernel rank approximation and sub-sample size, respectively.  The choice of $m_1$ and $m_2$ can be made based on computational feasibility or chosen via cross-validation.  

\begin{algorithm}[tbp]
\label{alg3}
\DontPrintSemicolon
\SetAlgoLined
\KwIn{kernel $k$, training data $\left\{ Y_{B(i)},  \{x_{i,l}\}_{l=1}^{r_i} \right\}_{i=1}^{n}$, and $m_1, m_2 \in \mathbb{N}$ s.t. $m_1 \leq m_2 \leq n$.}
\BlankLine
\Begin(create the Nystr\"{o}m feature map){
    Take $m_2$ stratified sub-samples from $\left\{ \{x_{i,l}\}_{l=1}^{r_i} \right\}_{i=1}^{n}$ as $\hat{x}_{1}, \dots, \hat{x}_{m_2}$
    \;
    $\hat{K} \leftarrow \left[k\left(\hat{x}_{i}, \hat{x}_{j}\right)\right]_{m_2 \times m_2}$\;
    Find eigenvalue, eigenvector pairs of $\hat{K}$ as $(\hat{\lambda}_j, \hat{\mathbf{v}}_j ), j = 1, \dots, m_2$ ranked by decreasing order of eigenvalues. 
    \;
    $\hat{V} \leftarrow \left( \hat{\mathbf{v}}_1, \dots, \hat{\mathbf{v}}_{m_1} \right)$ and $\hat{D} \leftarrow \textrm{diag}( \hat{\lambda}_1, \dots, \hat{\lambda}_{m_1} )$, using $m_1$ out of $m_2$ eigen-pairs. 
    \; 
    Return the feature map $\tilde{\phi}(x) = \hat{D}^{-1 / 2} \hat{V}^{\top}\left(k\left(x, \hat{x}_{1}\right), \ldots, k\left(x, \hat{x}_{m_2}\right)\right)^{\top}$\;
}
$z_i \leftarrow \frac{1}{r_i} \sum_{l = 1}^{r_i} \tilde{\phi}(x_{i,l})$, for $i=1,\dots,n$ 
\label{alg3-zi}
\;
\KwRet $(\boldsymbol{w}, b)$ from Problem \eqref{MI-SMM MIP} via an MIQP solver 
\;
\caption{\textsc{mi-smm~(miqp)}}
\end{algorithm}     

The advantage of Algorithm~\ref{alg3} is that it is guaranteed to find a global solution to \eqref{MI-SMM primal} under the feature map approximation. In contrast to Algorithm~\ref{alg2}, it is potentially much slower, as MIQP problems have exponential computational complexity and MIQP problems are NP-complete \citep{pia_mixed-integer_2017}. Despite this, solvers are improving their speed and can solve small problems efficiently through heuristic modifications of the branch and bound algorithm \citep{lazimy_mixed-integer_1982,gurobi_optimization_llc_mixed-integer_2021}.  

\section{Simulations} \label{simulations}

\subsection{Simulation data}

To evaluate our proposed approach, we consider four distinct simulation scenarios that contain data with weakly supervised labels and a distributional instance structure. Each scenario is parametrized by three quantities that control the size of the data: number of top-level groups (20, 50), number of instances per group (3, 6), and the sample size from each distributional instance (20, 50, 100), for 12 total combinations.  To construct the data, we first generate instance labels $y_i$ as i.i.d. Bernoulli random variables with $P(y_i =1) = 0.15$ for all instances.  Then the group label $Y_I$ is determined from the instance label through Assumption \ref{a:mil}.  The samples $\{ x_{i, j} \}_{j=1}^{r_i}$ are generated depending on the four distinct scenarios, where $\mathcal{X} = \mathbb{R}^{10}$, as described in Table \ref{tab:exp1-setup}.

\begin{sidewaystable}
\caption{\label{tab:exp1-setup}Summary of Simulations Setup}
\begin{threeparttable}
\begin{tabular}[t]{llll}
\toprule
\multicolumn{1}{c}{ } & \multicolumn{2}{c}{Differing Components} & \multicolumn{1}{c}{ }\\
\cmidrule(l{3pt}r{3pt}){2-3}
Setting                      & When $y_i = +1$                                                                                                                                              & When $y_i = -1$                                                                                                                                            & Remaining Components                                  \\
\midrule
Student $t$ vs. Normal       & $(x_{i,j}^{(1)}, \dots, x_{i,j}^{(5)}) \sim \textrm{MVT}\tnote{1} \left(\nu = 3, \boldsymbol{0}_5, \frac{1}{3} \boldsymbol{I}_5 \right)$                                   & $(x_{i,j}^{(1)}, \dots, x_{i,j}^{(5)}) \sim \textrm{MVN}\tnote{2} \left(\boldsymbol{0}_5,  \boldsymbol{I}_5 \right)$                                                     & $\textrm{MVN} \left(\boldsymbol{0}_5, \boldsymbol{I}_5 \right)$ \\
Covariance differences       & $(x_{i,j}^{(1)}, x_{i,j}^{(2)}) \sim \mathrm{MVN}\left(\mathbf{0}_{2},\left(\begin{smallmatrix}{} 1 & -0.5 \\ -0.5 & 1 \end{smallmatrix}\right)\right)$ & $(x_{i,j}^{(2)}, x_{i,j}^{(3)}) \sim \mathrm{MVN}\left(\mathbf{0}_{2},\left(\begin{smallmatrix}{} 1 & 0.5 \\ 0.5 & 1 \end{smallmatrix}\right)\right)$ & $\textrm{MVN} \left(\boldsymbol{0}_8, \boldsymbol{I}_8 \right)$ \\
Mean differences             & $(x_{i,j}^{(1)}, \dots, x_{i,j}^{(5)}) \sim \mathrm{MVN}\left(\mathbf{0.2}_{5},\boldsymbol{I}_5 \right)$                                                     & $(x_{i,j}^{(1)}, \dots, x_{i,j}^{(5)}) \sim \mathrm{MVN}\left(\mathbf{0}_{5},\boldsymbol{I}_5 \right)$                                                     & $\textrm{MVN} \left(\boldsymbol{0}_5, \boldsymbol{I}_5 \right)$ \\
Large covariance differences & $(x_{i,j}^{(1)}, \dots, x_{i,j}^{(5)}) \sim \mathrm{MVN}\left(\mathbf{0}_{5},\mathbf{\Sigma}_{S4} \right)$                                              & $(x_{i,j}^{(6)}, \dots, x_{i,j}^{(10)}) \sim \mathrm{MVN}\left(\mathbf{0}_{5},\mathbf{\Sigma}_{S4} \right)$                                           & $\textrm{MVN} \left(\boldsymbol{0}_5, \boldsymbol{I}_5 \right)$ \\
\bottomrule
\end{tabular}
\begin{tablenotes}
\footnotesize
We use $\boldsymbol{0}_p$ to denote a length $p$ vector with all elements being 0 (with similar notation for other scalars) and $\boldsymbol{I}_p$ to denote the $p \times p$ identity matrix.
\\
\item[1] $\textrm{MVT} \left(\nu, \boldsymbol{\delta}, \mathbf{\Sigma} \right)$ is the multivariate $t$ distribution with degrees of freedom $\nu \geq 3$, mean $\boldsymbol{\delta}$, covariance matrix $\frac{\nu}{\nu - 2} \mathbf{\Sigma}$,  and density $f(x; \nu, \boldsymbol{\delta}, \mathbf{\Sigma}) = \frac{\Gamma\left(\frac{v+k}{2}\right)}{\Gamma\left(\frac{v}{2}\right) \sqrt{(v \pi)^{k}|\mathbf{\Sigma}|}} \cdot\left(1+\frac{(x-\boldsymbol{\delta})^{t} \mathbf{\Sigma}^{-1}(x-\boldsymbol{\delta})}{v}\right)^{-\frac{v+k}{2}}$.
\\
\item[2] $\textrm{MVN} \left(\boldsymbol{\mu},  \mathbf{\sigma} \right)$ is the multivariate normal distribution with mean $\boldsymbol{\mu}$, covariance matrix $\mathbf{\Sigma}$, and density $f\left(x; \boldsymbol{\mu}, \mathbf{\Sigma}\right)=  \frac{1}{\sqrt{(2 \pi)^{k}|\mathbf{\Sigma}|}} \cdot \exp \left(-\frac{1}{2}(x-\boldsymbol{\mu})^{\mathrm{T}} \boldsymbol{\Sigma}^{-1}(x-\boldsymbol{\mu})\right)$.
\\
\item[3] $\mathbf{\Sigma}_{S4}$ is a $5 \times 5$ covariance matrix with 1 on the diagonal and 0.5 in every other entry. 
\end{tablenotes}
\end{threeparttable}
\end{sidewaystable}

\subsection{Methods compared} \label{sec:methods-compared}

Both algorithms for our proposed non-convex SVM solution, \textsc{mi-smm~(heuristic)} and \textsc{mi-smm~(miqp)}, were tested on the simulation scenarios and compared to two ad hoc approaches. The comparison methods simplify the data to a structure that can be solved by previous approaches.

\begin{longlist}
    \item \textsc{si-smm}: The first approach, which we call Single-Instance Support Measure Machines (\textsc{si-smm}), relies on imputing the spot label as the slide label $\tilde{y_i} = Y_I$. For the weakly-supervised setting, this results in the spot-level data set of $\{\tilde{y}_i, \mathbb{P}_i\}_{i=1}^n$, on which we can train an SMM classifier $h$ \citep{muandet_learning_2012}. A new slide $B_{I'} = \{ \mathbb{P}_i : i \in I'\}$ can be classified as $\hat{Y}_{I'} = \max_{i \in I'} h(\mathbb{P}_i)$, following Assumption \ref{a:mil}. This technique has previously been considered for data with the weakly supervised label structure, but not for distributional instance data \citep{ray_supervised_2005,alpaydin_single-_2015}.  
    
    \item \textsc{mi-svm}: The second approach focuses on summarizing each distributional instance into a feature vector.  We define a function $\varphi : \mathcal{P} \to \mathbb{R}^p$ which summarizes a distribution into a feature vector: $\varphi(\mathbb{P}_i) = z_i$.  For example, this could include the mean, standard deviation, or correlation of $\mathbb{P}_i$. Applying $\varphi$ results in a group of feature vectors $B_{I}^\varphi = \{ z_i : i \in I\}$ corresponding to each original slide-level group $B_I$.  We apply a similar non-convex SVM algorithm to the data set $\{Y_I, B_{I}^\varphi\}$, called \textsc{mi-svm} from  \cite{andrews_support_2003}. 
   
    For this approach, we consider three sets of summary functions on the instance distributions $\hat{\mathbb{P}}_i = \{x_{i,j}\}_{j=1}^{r_i}$ where $x_{i,j} \in \mathcal{X} \subseteq \mathbb{R}^d$:
    \begin{itemize}
        \item $\varphi_{\textrm{univ}1} = \left[\textrm{mean}(\hat{\mathbb{P}}_i), \textrm{SD}(\hat{\mathbb{P}}_i)\right] \in \mathbb{R}^{2d}$, which includes the sample mean and standard deviation of $\mathbb{P}_i$. 
        
        \item $\varphi_{\textrm{univ}2} = \left[\textrm{skew}(\hat{\mathbb{P}}_i), \textrm{kurt}(\hat{\mathbb{P}}_i), \textrm{Q1}(\hat{\mathbb{P}}_i), \textrm{Q3}(\hat{\mathbb{P}}_i) \right] \in \mathbb{R}^{4d}$, which includes the sample skew, sample kurtosis, and the 25th and 75th percentile of $\mathbb{P}_i$.
        
        \item $\varphi_{\textrm{cor}}$: 2-way correlations.  All 2-way correlations between each pair of features are computed using the sample Pearson correlation coefficient, so that $\varphi_{\textrm{cor}}(\hat{\mathbb{P}}_i) \in \mathbb{R}^{p}$ where $p = \binom{d}{2}$.
    \end{itemize}
    From these three sets of summary statistics, we evaluate four distinct combinations.  We name each combination based on the concatenated set of included functions: \textsc{mi-svm (univ1)}, \textsc{mi-svm (univ1, univ2)}, \textsc{mi-svm (univ1, cor)}, \textsc{mi-svm (univ1, univ2, cor)}.  All four utilize the first set of univariate features $\varphi_{\textrm{univ}1}$, while the second set of univariate features $\varphi_{\textrm{univ}2}$ and the multivariate features $\varphi_{\textrm{cor}}$ are toggled on or off. 

\end{longlist}

We use the Gaussian kernel $k(x,z) = \exp \left( - \frac{\left\lVert x - z \right\rVert^2}{2 \sigma^2} \right)$ in \textsc{si-smm} and \textsc{mi-smm} as the embedding kernel, and in all \textsc{mi-svm} approaches to introduce non-linearity.  

\subsection{Implementation details}

Each combination of data size and data scenario is evaluated similarly.  We generate 100 data sets to serve as training data.  For training, we employ a 5-fold cross-validated grid-search to determine the optimal choice of $C$ and $\sigma$ in the SVM optimization problem and Gaussian kernel, respectively. The parameter $C$ is weighted inversely proportional to the number of positive groups and the number of negative instances to help alleviate potential class imbalance.  A corresponding testing data set of 500 groups is generated for each training data set.  We evaluate model performance on the predicted top-level group labels from the area under the receiver operating characteristic curve (AUROC).

On each data set, we apply the seven distinct methods described previously.  Since there are 10 covariates for each data point, the \textsc{mi-svm} methods have 20, 60, 65, and 105 total features for \textsc{uinv1}, \textsc{uinv1+uinv2}, \textsc{uinv1+cor}, \textsc{uinv1+uinv2+cor}, respectively.  For the \textsc{mi-smm (miqp)}, we take $m_1 = 240$ samples and use rank $m_2 = 240$ for the Nystr\"{o}m kernel approximation. 

\subsection{Performance evaluation}

\begin{figure}[tbp]
    \centering\includegraphics[width=0.99\linewidth]{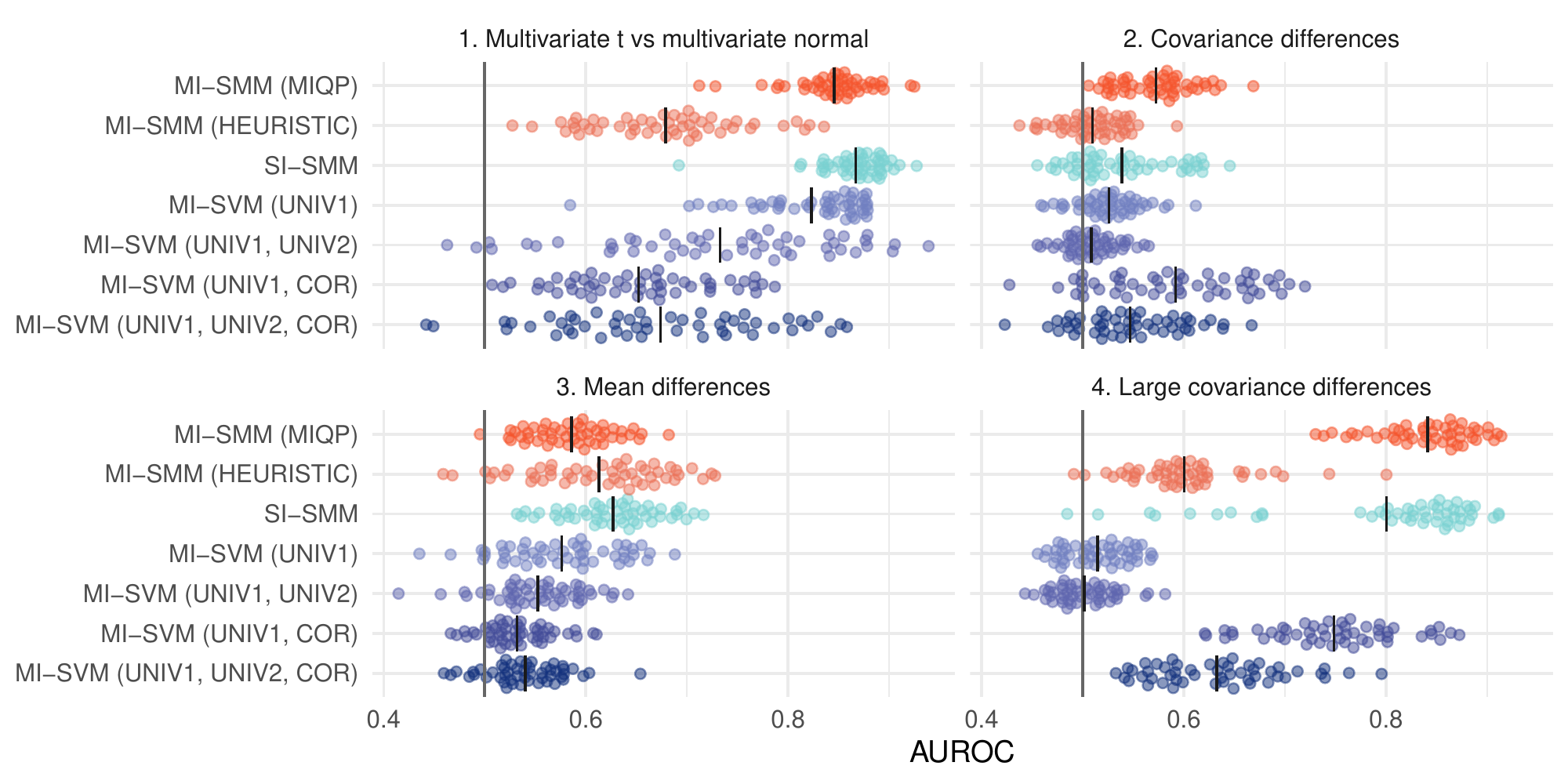}
    \caption{Scatter plot of AUROC vs method for four simulation scenarios when there are 100 top-level groups, 3 instance per group, and 50 distribution samples per instance. The points in each subplot represent performance on 100 individually simulated combinations of training and testing data sets.  The width of each cluster of points closely corresponds to the number of points with the given performance. A black line shows the mean AUROC within each scenario and method.}
    \label{fig:four-scenarios}
\end{figure}

Figure \ref{fig:four-scenarios} compares the AUROC performance of each method for the data sets generated with 50 groups, 3 instances per group, and 50 samples from each instance distribution.  Overall, \textsc{mi-smm (miqp)} and \textsc{si-smm} show strong performance in most scenarios.  In Scenario 3, where the distributions differ in their means, \textsc{si-smm} outperforms \textsc{mi-smm (miqp)} by a fair margin, but it is less consistent in Scenarios 2 and 4.    

\begin{figure}[tbp]
    \centering\includegraphics[width=0.99\linewidth]{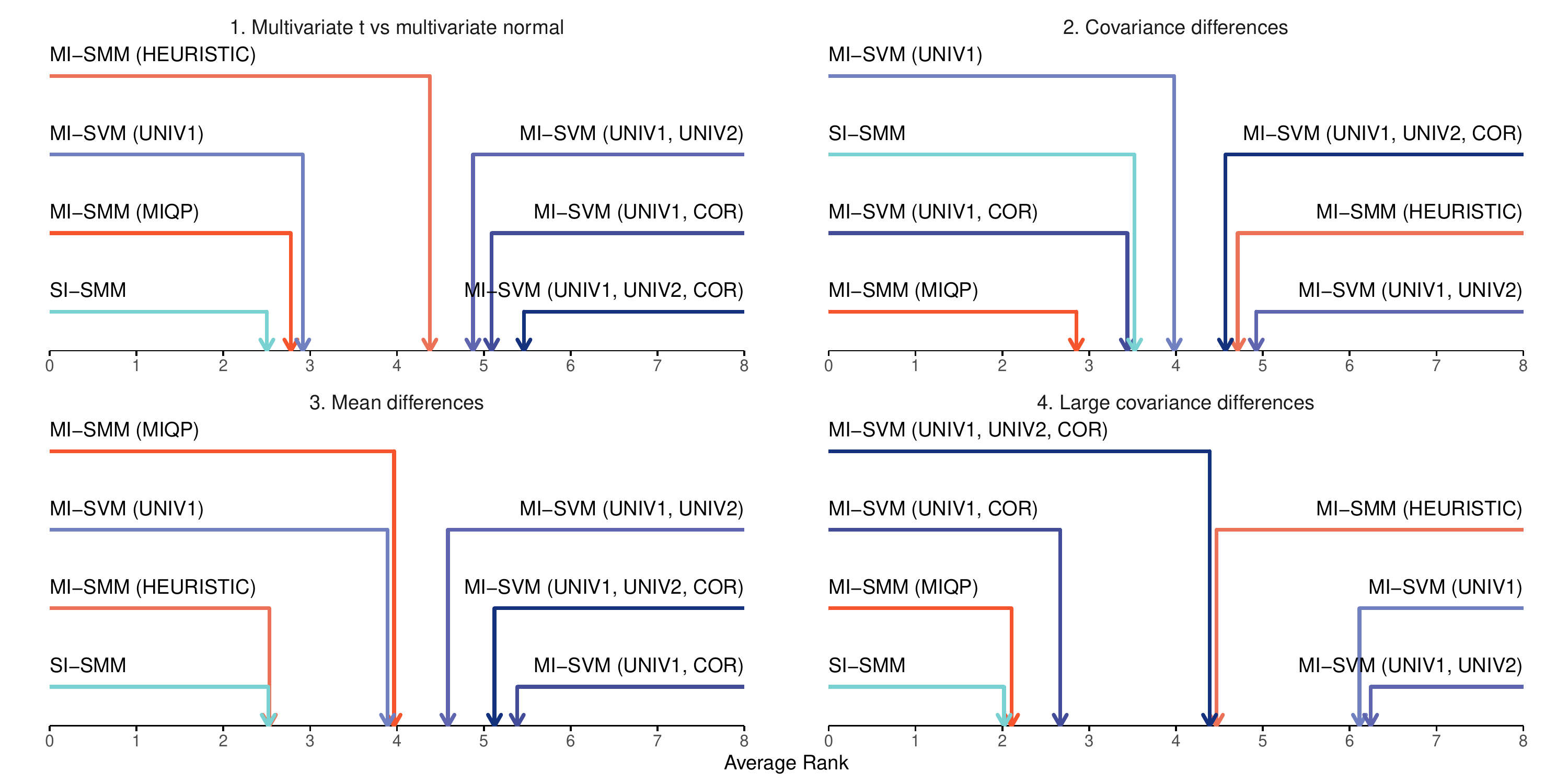}
    \caption{Average ranks of test AUROC across all data sets. Each panel represents one simulation scenario. Ranks range from 1-7, with a rank close to 1 indicating the best performance.}
    \label{fig:avg-rank}
\end{figure}

Figure~\ref{fig:avg-rank} summarizes the results across the 1,200 data sets (12 sample size combinations $\times$ 100 data sets) for each of the simulation scenarios by providing the average classifier AUROC rank (1-7). The trends illustrated are similar to those in Figure \ref{fig:four-scenarios}. The performance of the \textsc{mi-svm} methods depends heavily on the scenario.  They perform better when the summary features correspond to the scenario differences.  However, it's not sufficient to throw in all features since \textsc{mi-svm (univ1, univ2, cor)} is not the strongest performing of the \textsc{mi-svm} methods in any data scenario. The full results, including plots of AUROC for each scenario and sample size combination, are available in the Supplementary Materials. 

\section{Analysis of collagen features data} \label{analysis}

Using our non-convex SVM methods, \textsc{mi-smm}, proposed in Section~\ref{methods}, we analyze the collagen features data. This section includes an overview of the key variables and an evaluation of how well our methodology predicts the tumor label compared to other approaches.

\subsection{Data set overview}

244 DCIS patients have a total of 348 tissue sample slides, which represent the top-level groups.  We seek to classify the 235 tumor tissue slides ($Y_I = 1$) vs. the 113 normal tissue slides ($Y_I = -1$).  Instances are represented through the multiple spots chosen on each image (min 1, max 8), as shown in Figure~\ref{fig:hists}. These spots yield a weakly-supervised label data structure. We assume that tumorous tissue samples have at least one spot with features indicating tumor, while non-tumorous samples have no spots indicating tumor, which aligns with Assumption~\ref{a:mil}. 

Each spot was processed using the second harmonic generating (SHG) imaging \citep{chen_second_2012} to find morphological features on numerous collagen fibers (min 73, max 1092).  Since there are many collagen fibers within each spot, and the relationship between the fibers may be predictive of tumor tissue, we consider each spot as an empirical probability distribution, where fibers are sampled from the spot. Figure~\ref{fig:hists} shows that the number of fibers is commonly between 350 -- 750.

This SHG process creates 20 features such as length, width, curvature, and measurements of alignment and density. Table~\ref{tab:fiber-comparison} provides an overview of all features and compares the mean of subject-level averages between patients having tumorous and non-tumorous slides. Many features show significant differences in the two groups, most notably the distance of one fiber to other nearby fibers, the angle to the nearest relative boundary, and the box alignment. Some features are heavily skewed, which we log-transform. All are centered and scaled in training.

\begin{table}

\caption{\label{tab:fiber-comparison}Comparison of subject-level averages of morphologic features.}
\centering
\begin{tabular}[tb]{lrrr}
\toprule
\multicolumn{1}{c}{ } & \multicolumn{2}{c}{Mean} & \multicolumn{1}{c}{ } \\
\cmidrule(l{3pt}r{3pt}){2-3}
Variable & Non-tumor & Tumor & P value\\
\midrule
Total length & 57.97 & 58.63 & 0.0570\\
End-to-end length & 52.80 & 53.51 & 0.0345\\
Width & 5.55 & 5.46 & 0.0023\\
Nearest relative boundary angle & 8.33 & 10.44 & <0.0001\\
Curvature & 0.91 & 0.92 & 0.0783\\
Distance to nearest 2 & 24.30 & 25.14 & <0.0001\\
Distance to nearest 4 & 31.33 & 32.55 & <0.0001\\
Distance to nearest 8 & 42.00 & 43.91 & <0.0001\\
Distance to nearest 16 & 58.13 & 61.30 & <0.0001\\
Mean nearest distance & 38.94 & 40.73 & <0.0001\\
Std nearest distance & 14.80 & 15.83 & <0.0001\\
Alignment of nearest 2 & 0.76 & 0.77 & 0.0077\\
Alignment of nearest 4 & 0.64 & 0.65 & 0.0014\\
Alignment of nearest 8 & 0.56 & 0.58 & 0.0015\\
Alignment of nearest 16 & 0.51 & 0.53 & 0.0069\\
Mean nearest alignment & 0.62 & 0.63 & 0.0031\\
Std nearest alignment & 0.17 & 0.17 & 0.0035\\
Box alignment 32 & 0.87 & 0.88 & 0.0086\\
Box alignment 64 & 0.68 & 0.70 & <0.0001\\
Box alignment 128 & 0.54 & 0.57 & <0.0001\\
\bottomrule
\end{tabular}
\end{table}

\subsection{Predictive performance}

We compare the predictive performance of our proposed non-convex SVM framework to the comparison methods from Section~\ref{sec:methods-compared}. We use 10 replications of a 10-fold cross-validation procedure to evaluate each method \citep{kim_estimating_2009}.  In each training fold, a 5-fold cross-validated grid-search determines the optimal choice of $C$ and $\sigma$ in the SVM optimization problem and Gaussian kernel, respectively.  Test performance is combined across the testing folds through averaging.  For the Nystr\"{o}m approximation in  \textsc{mi-smm (miqp)}, we take $m_1 =$ 360 samples and use rank $m_2 =$ 360.  Due to the data set size, the \textsc{mi-smm (miqp)} algorithm takes considerable time with state-of-the-art optimization solvers \citep{gurobi_optimization_llc_mixed-integer_2021}; we accept the best solution after 10 minutes.

\begin{table}

\caption{\label{tab:fiber-results}Summary of predictive performance.}
\centering
\begin{tabular}[tb]{lrr}
\toprule
\multicolumn{1}{c}{ } & \multicolumn{2}{c}{AUROC} \\
\cmidrule(l{3pt}r{3pt}){2-3}
Method & Mean & SD\\
\midrule
MI-SMM (MIQP) & \textbf{0.768} & \textbf{0.023}\\
MI-SMM (HEURISTIC) & 0.695 & 0.017\\
SI-SMM & 0.695 & 0.025\\
MI-SVM (UNIV1) & 0.761 & 0.013\\
MI-SVM (UNIV1, UNIV2) & \textbf{0.772} & \textbf{0.016}\\
MI-SVM (UNIV1, COR) & \textbf{0.765} & \textbf{0.015}\\
MI-SVM (UNIV1, UNIV2, COR) & 0.759 & 0.018\\
\bottomrule
\end{tabular}
\end{table}

The AUROC performance for each method on the fiber-feature data set is summarized in Table \ref{tab:fiber-results}.  The top methods on this data are \textsc{mi-svm} (univ1, univ2), \textsc{mi-smm (miqp)}, and \textsc{mi-svm} (univ1, cor), although all of the \textsc{mi-svm} methods do well with an AUROC near 0.77. The mean and standard deviation of features from $\varphi_{\textrm{univ}1}$ are adequate to predict tumor or normal tissue; however, this performance can be slightly improved with additional univariate and multivariate features.  Importantly, \textsc{mi-smm (miqp)} is comparable with these methods without requiring a potentially arbitrary choice in these features.  



\section{Discussion} \label{discussion}

\textsc{mi-smm}, a non-convex, maximum-margin-based learning approach, was constructed to classify tumorous tissue vs. non-tumorous tissue from a data set of collagen fiber features within the tumor microenvironment. It directly handles both aspects that make the data set unique: the weakly-supervised label structure stemming from the multiple spots per subject and the distributional instance structure stemming from many fibers per spot. In addition, \textsc{mi-smm} allows for non-linear transformations of the data through an embedding kernel and two separate algorithms that trade off computational efficiency for optimality guarantees. In its entirety, it offers flexibility to adapt to similarly-structured, unseen data sets.

Our approach shows strong performance across all of the experimental settings we tested in this work. The other ad-hoc approaches worked well in either the experimental simulations or the motivating data set, but not both, showing that they were more data-dependent. In contrast, \textsc{mi-smm} was more adaptable to the data sets we tested on. Further testing on additional data is needed to make broad conclusions, but this shows a promising start. 

This work outlines the first combination of weakly-supervised labels and distributional instances that we are aware of, but data sets with similar structures may likely surface in the future. Thinking of instances as distributions is justified when we observe a large sample of random events from a unique identity. As the size and number of data sets grow in the future, examples of this structure may become more common. Such examples could include: the behavior of a single user on an internet platform across many days; weights from trained inner layers of a neural network; or images under a sample of scaling, translation, and rotation transformations  \citep[see][]{muandet_learning_2012}. The weakly-supervised setting has already occurred in previous works \citep{dietterich_solving_1997,andrews_support_2003,ray_supervised_2005,campanella_clinical-grade_2019,chen_lung_2022}. It spans a host of applications, such as drug discovery, tumor classification from whole slide images, and text document classification. We hope future researchers can use our methodology on data with both distributional instances and weakly-supervised label structure.

One limitation of the motivating data set is the assumption of how tissue slide labels relate to the spot labels. If, for a given tumorous tissue sample, no spots are observed that contain the features of cancer, this could violate our assumption and provide incorrect label information to the training algorithm. These violations are unlikely in practice because of the expert selection of significant spots. Another limiting factor is the computation in large data sets. While \textsc{mi-smm (miqp)} works better than \textsc{mi-smm (heuristic)}, the computational cost of solving the MIQP problem directly is high, as the number of integer combinations multiplies rapidly with an increasing number of instances. Setting a time limit on this algorithmic approach worked in practice on the collagen fiber data set, despite the large optimality gap observed. Alternatively, the heuristic algorithm for \textsc{mi-smm} is typically quite fast, even for large data sets.  

All approaches described in this work are available in an open-source R package, \verb|mildsvm|.\footnote{Available at \href{https://github.com/skent259/mildsvm}{github.com/skent259/mildsvm}} The package is designed for building models with the familiar formula interface or through data frames. It also includes useful tools for SVM-based methods, including methods for the kernel feature approximation of Section~\ref{sec:fm-approx}, calculation of kernels on probability distributions \citep{muandet_learning_2012}, and summarization of distributional instances to features. We believe this work can provide a basis to expand the literature on non-convex SVMs, potentially through additional kernels that work on probability distribution data or alternative non-convex SVM problems targeting different assumptions of data structure.

\begin{funding}
Research reported in this publication was partially supported by the National Cancer Institute of the National Institutes of Health under Award Number T32LM012413 and the University of Wisconsin Carbone Cancer Center Support Grant P30 CA014520. The content is solely the responsibility of the authors and does not necessarily represent the official views of the National Institutes of Health.
\end{funding}

\begin{supplement}
\stitle{Supplement to ``Non-convex SVM for cancer diagnosis based on morphologic features of tumor microenvironment''}
\sdescription{The supplementary material contains proofs of all theoretical results and additional simulation results (pdf file). An associated R package mildsvm is currently available on GitHub (\href{https://github.com/skent259/mildsvm}{github.com/skent259/mildsvm}) and will be made available on the Comprehensive R Archive Network (CRAN). Full scripts to recreate the simulation results and data analysis are provided in the following GitHub repository for reproducibility: \href{https://github.com/skent259/mildsvm-sims}{github.com/skent259/mildsvm-sims}; the motivating data set is excluded from the public repository but can be made available upon request.}
\end{supplement}


\bibliographystyle{references/imsart-nameyear} 
\bibliography{references/biblio_aoas}       

\end{document}


\begin{frontmatter}
\title{Supplement to ``Non-convex SVM for cancer diagnosis based on morphologic features of tumor microenvironment''}
\runtitle{Supplement: Non-convex SVM on tumor microenvironment features}

\begin{aug}
\author[A]{\fnms{Sean}~\snm{Kent}\orcid{0000-0001-8697-9069}}
\and
\author[B]{\fnms{Menggang}~\snm{Yu}\ead[label=e2]{meyu@biostat.wisc.edu}\orcid{0000-0002-7904-3155}}

\address[A]{Department of Statistics,
University of Wisconsin -- Madison}

\address[B]{Department of Biostatistics and Medical Informatics,
University of Wisconsin -- Madison\printead[presep={,\ }]{e2}}
\end{aug}




\end{frontmatter}

\section{Proofs of theoretical results} \label{proofs}

\subsection{Optimization trick for turning max constraints to mixed-integer constraint: Lemma and Proof}

\begin{lemma*}
    \label{lemma-opt}
    Given scalars $\{a_i\}_{i=1}^n, b \in \mathbb{R}$, there exists $L > 0$ sufficiently large so that the constraint $\max_{i} a_i \geq b$ is equivalent to
    \begin{equation}
    \label{lem2-star}
        \begin{aligned}
            & a_i \geq b - L \cdot z_i, &\forall i 
            \\
            & \zeta_{i} \in \{0, 1\}, &\forall i
            \\
            & \sum_{i=1}^n \zeta_{i} \leq n - 1, 
        \end{aligned}
    \end{equation}
\end{lemma*}

\begin{proof}
    Choose any $L$ such that $L \geq \max_i (b - a_i) = b - \min_i a_i$ and $L > 0$.  Let $j = \arg\max_i a_i$.

    Suppose $\max_{i} a_i \geq b$. By the choice of $L$, we have
    \begin{equation*}
        \begin{aligned}
        & 0 \geq b - a_j 
        \\
        & L \geq b - a_i & \text{for } i \neq j
        \end{aligned}
    \end{equation*}
    
    If $\zeta_j = 0$ and $\zeta_i = 1$ for $i \neq j$, then $L \cdot \zeta_i \geq b - a_i$ $\forall i$ and all constraints in Equation \eqref{lem2-star} are satisfied.
    
    Suppose, on the other hand, that Equation \eqref{lem2-star} holds.  Then the last line of \eqref{lem2-star} implies there exists $j$ such that $\zeta_j = 0$.  Thus, from the first line of \eqref{lem2-star}, we have $a_j \geq b$ and so 
    \begin{equation*}
        \max_i a_i \geq a_j \geq b     
    \end{equation*}
\end{proof}

\subsection{Proof of Lemma 3.1 on feature map approximation from distributional instances}

\begin{customthm}{3.1}
\label{lemma-fm}
    Suppose that a kernel function $k : X \times X \to \mathbb{R}$ from a RKHS admits a feature map $\phi : X \to F \subset \mathbb{R}^p$ so that $k(x,z) = \langle \phi(x), \phi(z) \rangle$.  Then 
    \begin{equation*}
        K_{\textrm{emp}} (\hat{\mathbb{P}}_i, \hat{\mathbb{P}}_j) = \langle \bar{\phi}(\hat{\mathbb{P}}_i), \bar{\phi}(\hat{\mathbb{P}}_j) \rangle
    \end{equation*}
    where $\hat{\mathbb{P}}_i$ and $\hat{\mathbb{P}}_j$ are empirical distributions of $\mathbb{P}_i$ and $\mathbb{P}_j$ given random samples $\{x_{i,l} \}_{l=1}^{r_i}$ and $\{x_{j,m} \}_{m=1}^{r_j}$, and $\bar{\phi}(\hat{\mathbb{P}}_i) = \frac{1}{r_i} \sum_{l = 1}^{r_i} \phi(x_{i,l})$ and $\bar{\phi}(\hat{\mathbb{P}}_j) = \frac{1}{r_j} \sum_{m = 1}^{r_j} \phi(x_{j,m})$ are the empirical means of the the random samples evaluated on the feature map.  
\end{customthm}

\begin{proof}
    The result follows from direct calculation and reordering of the summations: 
    \begin{align*}
        K_{\textrm{emp}} (\hat{\mathbb{P}}_i, \hat{\mathbb{P}}_j) 
        &= \frac{1}{r_i \cdot r_j} \sum_{l = 1}^{r_i} \sum_{m = 1}^{r_j} k(x_{i,l}, x_{j,m}) 
        \\
        &= \frac{1}{r_i \cdot r_j} \sum_{l = 1}^{r_i} \sum_{m = 1}^{r_j} \langle \phi(x_{i,l}),  \phi(x_{j,m}) \rangle
        \\
        &= \frac{1}{r_i \cdot r_j} \sum_{l = 1}^{r_i} \sum_{m = 1}^{r_j} \sum_{k = 1}^p \phi_k(x_{i,l})  \phi_k(x_{j,m})
        \\
        &= \sum_{k = 1}^p \left( \frac{1}{r_i} \sum_{l = 1}^{r_i} \phi_k(x_{i,l}) \right) \left( \frac{1}{r_j} \sum_{m = 1}^{r_j} \phi_k(x_{j,m})  \right)
        \\
        &= \langle \bar{\phi}(\hat{\mathbb{P}}_i), \bar{\phi}(\hat{\mathbb{P}}_j \rangle
    \end{align*}
    where $\phi_k(x)$ is the $k$th component of the vector $\phi(x) \in \mathbb{R}^p$. 
\end{proof}

\section{Additional Experimental Results} \label{more-experiments}

We present additional results for the experiments considered in Section 4 of the main article. 
The following plots display the AURUC of each comparison method across the 100 simulated data sets.  Each panel represents a combination of the data size, and each plot represents a different simulation scenario. 

\begin{figure}[!p]
    \centering\includegraphics[width=0.99\linewidth]{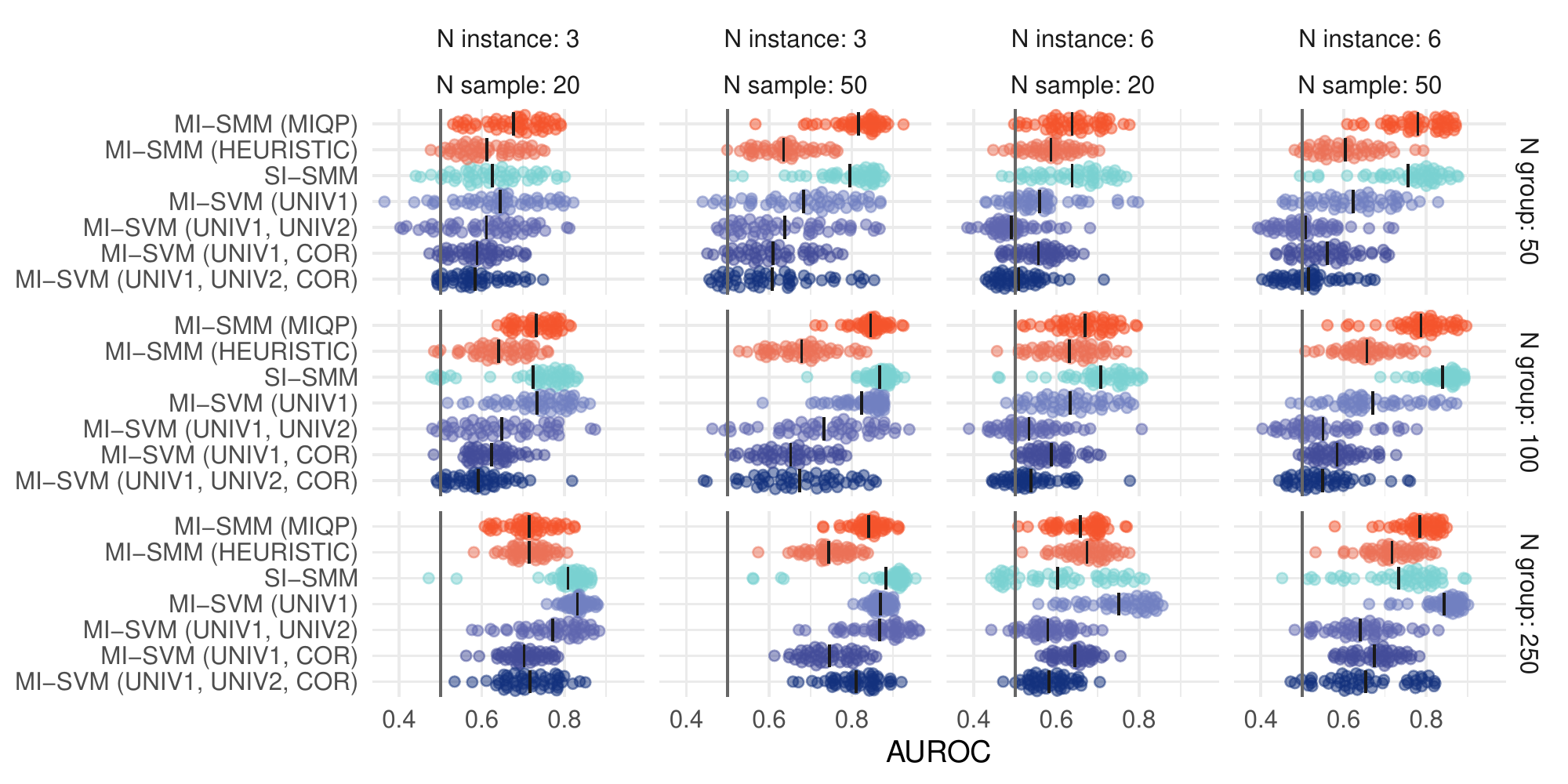}
    \caption{Scatter plot of AUROC vs method for Student \textit{t} vs Normal simulations. Each panel contains a separate data size scenario. The points in each subplot represent performance on 100 individually simulated combinations of training and testing data sets.  The width of each cluster of points closely corresponds to the number of points with the given performance. A solid black line shows the mean AUROC within each scenario and method.}
    \label{fig:appendix-1}
\end{figure}

\begin{figure}[!p]
    \centering\includegraphics[width=0.99\linewidth]{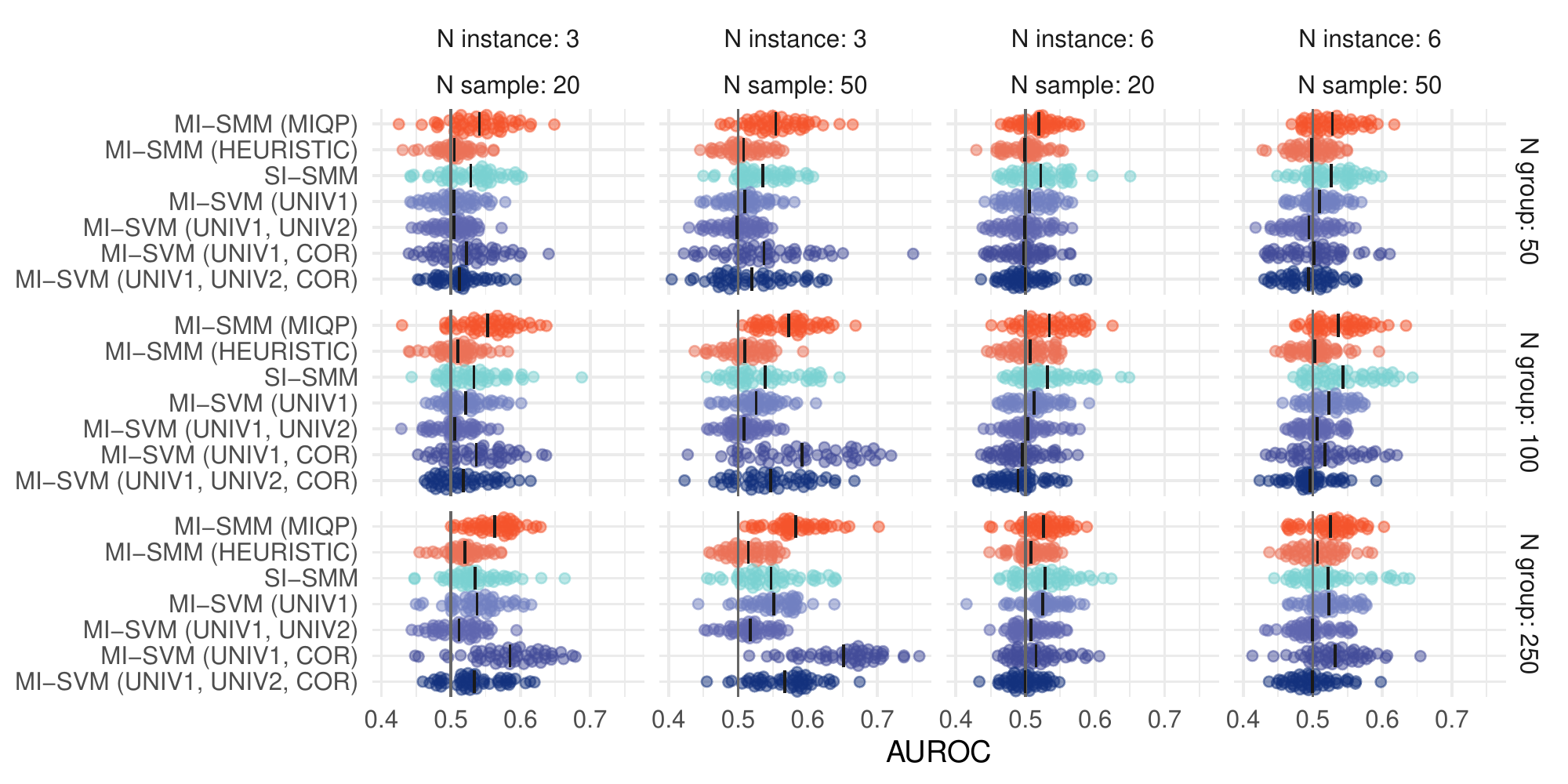}
    \caption{Scatter plot of AUROC vs method for the covariate differences simulations. Each panel contains a separate data size scenario. The points in each subplot represent performance on 100 individually simulated combinations of training and testing data sets.  The width of each cluster of points closely corresponds to the number of points with the given performance. A solid black line shows the mean AUROC within each scenario and method.}
    \label{fig:appendix-2}
\end{figure}

\begin{figure}[!p]
    \centering\includegraphics[width=0.99\linewidth]{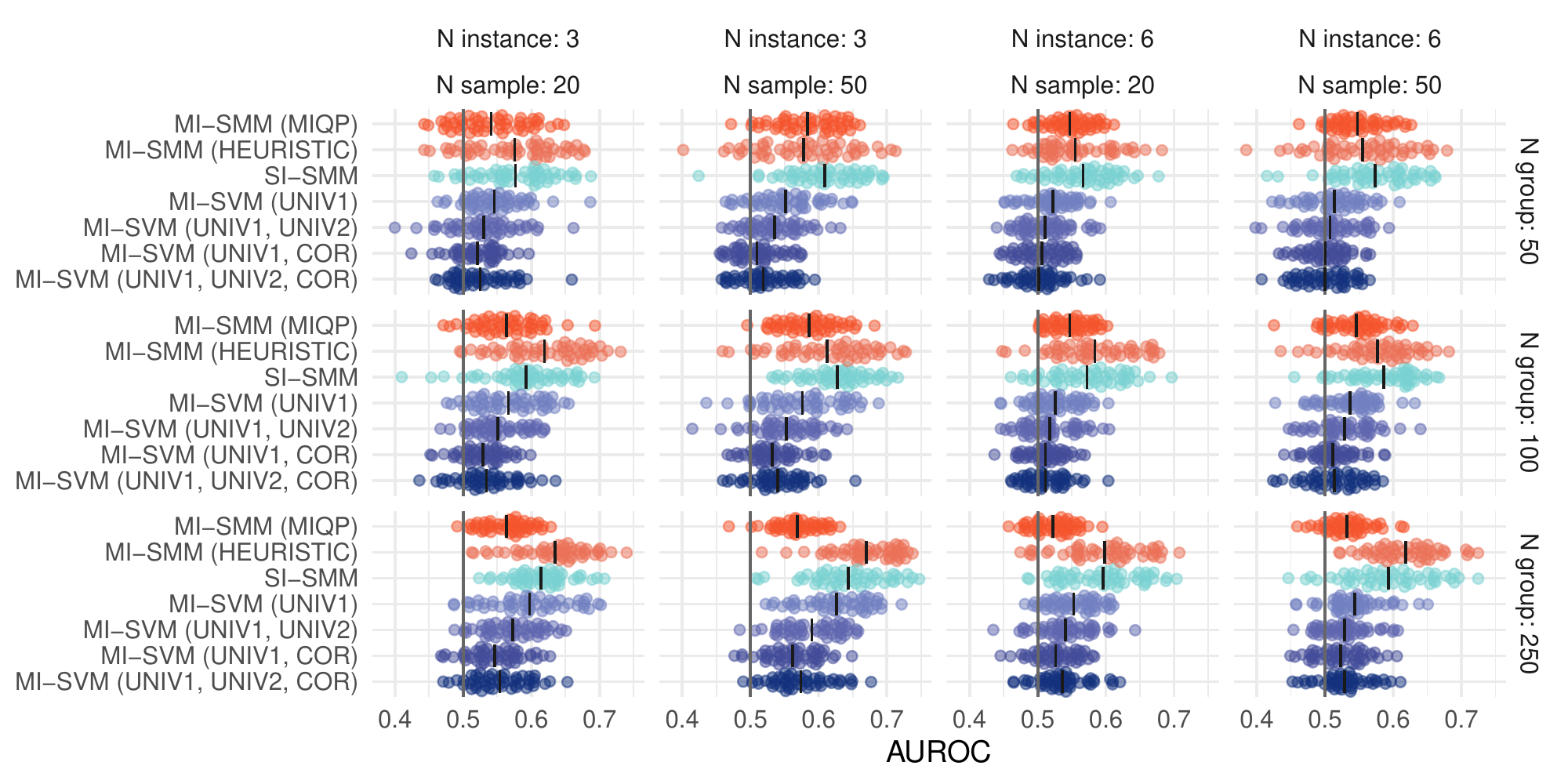}
    \caption{Scatter plot of AUROC vs method for the mean differences simulations. Each panel contains a separate data size scenario. The points in each subplot represent performance on 100 individually simulated combinations of training and testing data sets.  The width of each cluster of points closely corresponds to the number of points with the given performance. A solid black line shows the mean AUROC within each scenario and method.}
    \label{fig:appendix-3}
\end{figure}

\begin{figure}[!p]
    \centering\includegraphics[width=0.99\linewidth]{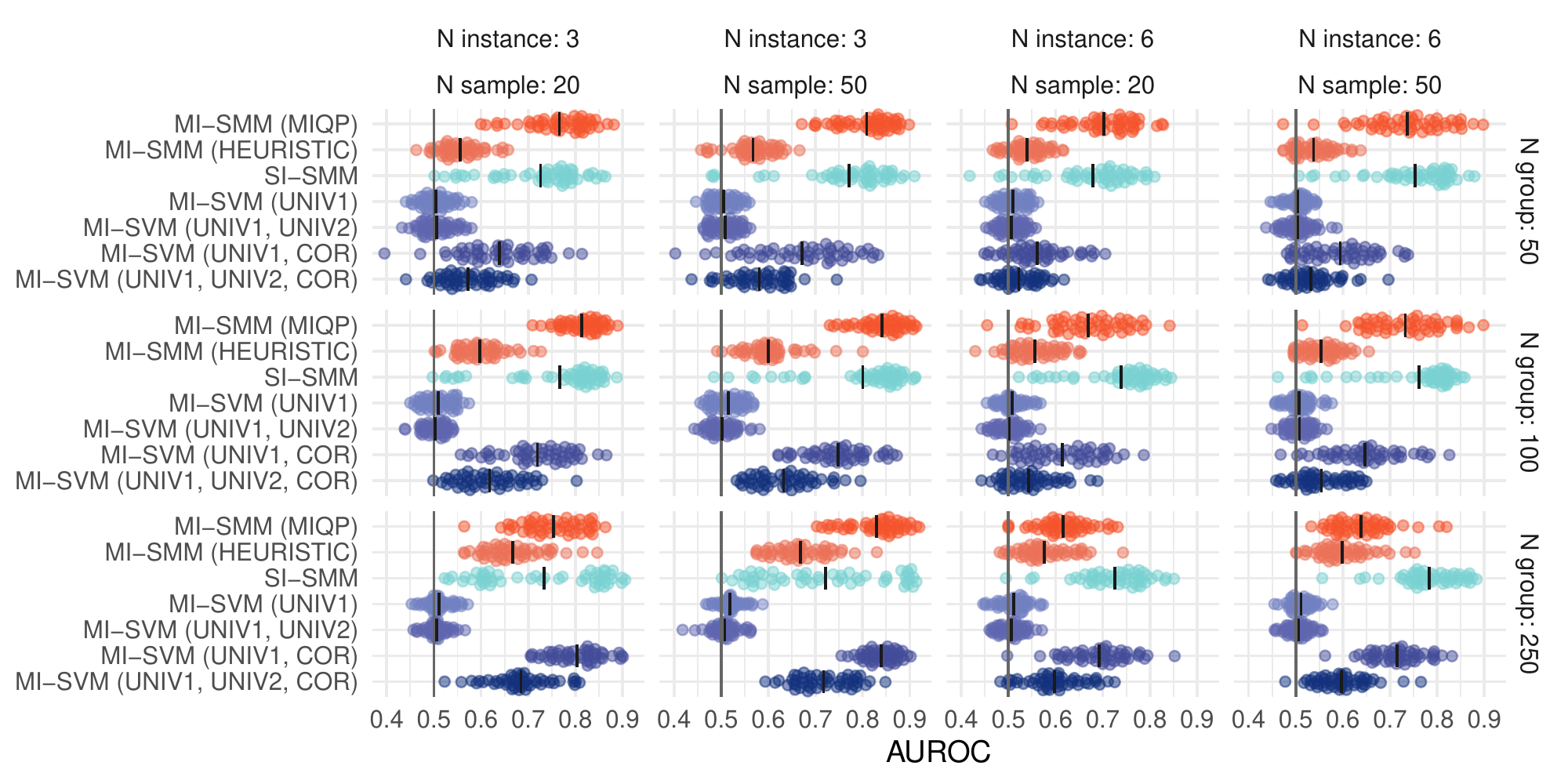}
    \caption{Scatter plot of AUROC vs method for the large covariance differences simulations. Each panel contains a separate data size scenario. The points in each subplot represent performance on 100 individually simulated combinations of training and testing data sets.  The width of each cluster of points closely corresponds to the number of points with the given performance. A solid black line shows the mean AUROC within each scenario and method.}
    \label{fig:appendix-4}
\end{figure}